# Theoretical Insights into 1:2 and 1:3 Internal Resonance for Frequency Stabilization in Nonlinear Micromechanical Resonators

Ata Donmez, Hansaja Herath, Hanna Cho*

Department of Mechanical and Aerospace Engineering
The Ohio State University
Columbus, OH 43210

**Abstract**

Micromechanical resonators are essential components in time-keeping and sensing devices due to their high frequency, high quality factor, and sensitivity. However, their extremely low damping can lead to various nonlinear phenomena that can compromise frequency stability. A major limiting factor is the Duffing hardening effect, which causes frequency drift through amplitude variations, known as the amplitude-frequency effect. Recently, internal resonance (InRes) has emerged as an effective approach to mitigate this issue and enhance frequency stabilization. In this study, we investigate the frequency stabilization mechanisms of 1:2 and 1:3 InRes using a generalized two-mode reduced-order model that includes Duffing nonlinearity and nonlinear modal coupling. By analyzing the frequency response curves and $\pi/2$–backbone curves, we demonstrate how different parameters affect the effectiveness of frequency stabilization. Our results identify two distinct regimes depending on the coupling strength relative to the stiffening effect as a key factor in determining the stabilization mechanism. For the regime of weak coupling, both 1:2 and 1:3 InRes achieve frequency stabilization through amplitude and frequency saturation over a range of forcing amplitudes. In contrast, strong coupling reduces the amplitude-frequency effect by forming an asymptote line for 1:2 InRes or a zero-dispersion point for 1:3 InRes. These insights offer valuable guidelines for designing micromechanical resonators with high-frequency stability, highlighting InRes as a robust tool for enhancing performance in practical applications.

(*) Corresponding Author, email: cho.867@osu.edu

# 1 Introduction

This paper presents an in-depth theoretical investigation into the frequency stabilization effect of internal resonance (InRes) in resonant micromechanical systems. These devices play a vital role in various sensors [1,2] and time-keeping applications [3,4] within micro/nano-electro-mechanical systems (MEMS/NEMS). For these applications, increasing the resonant amplitude and quality factor is often desired to enhance the signal strength and performance. However, this often pushes the resonators out of linear regimes, leading them to exhibit a rich spectrum of nonlinear phenomena. These nonlinear behaviors include hardening and/or softening nonlinear resonances [5,6], parametric resonances [7], sub- and super-harmonic resonances [8], and internal modal couplings between different vibratory modes with commensurate frequency ratios [9–11]. The source of these nonlinearities is typically rooted in geometric effects due to large deformations of the microstructures and electrostatic actuation [12–14]. As a result, these nonlinear effects alter the resonant response, making the natural frequency ($\omega$) of the resonator dependent on its vibrational energy $E$, i.e., $\omega(E)$, known as the amplitude-frequency effect [15,16]. From a frequency stability standpoint, this nonlinear behavior is a major limiting factor, as any variations in amplitude lead to corresponding frequency fluctuations [6,17,18]. This effect poses a substantial challenge in practical applications where precise frequency control is essential [19,20].

In recent years, several studies investigated a special case where the frequency $\omega(E)$ is not a monotonic function of energy but instead features an extremum, $d\omega/dE = 0$, or "zero-dispersion" point [6,15,18,21–23]. At this point, opposing nonlinearities (hardening and softening) are balanced, yielding zero-dispersion of frequency and reducing the amplitude-frequency effect in the vicinity of the extremum point. Various mechanisms have been explored to achieve zero-dispersion in micromechanical resonators. For instance, Refs [16,24,25] utilized the softening nonlinearity from electrostatic actuation to counteract the mechanical hardening effect within the microbeam. Rosenberg and Shoshani proposed using the initial curvature of an arch microbeam can introduce the necessary softening nonlinearity to be balanced with the hardening effect during resonance [6]. While these approaches have some promise, they also present design challenges, such as the need for precise geometric control and sensitivity to fabrication inconsistencies.

More recently, InRes has emerged as an effective way to stabilize the frequency of resonant MEMS [15,18,20,22,23]. When InRes is triggered, nonlinear coupling terms facilitate energy exchange between vibratory modes with commensurate frequencies ($n\omega_1 = \omega_2$ for 1:$n$ InRes, where $n$ is and integer and $\omega_i$ is the natural frequency of $i$-th mode). Early experimental studies of 1:3 InRes in open-loop operation demonstrated saturation in jump-down frequency at $\Omega \approx \omega_2/3$ as excitation amplitude is increased, where $\Omega$ is the excitation frequency [18,26]. Detailed theoretical analysis further suggested that nonlinear coupling introduces softening nonlinearity when excitation frequencies $\Omega < \omega_2/3$. With this, two competing effects, softening from intermodal coupling and hardening effects from mid-plane stretch, create zero-dispersion of frequency near $\Omega \approx \omega_2/3$ [15]. Other studies investigated whether similar frequency stabilization can be achieved through 1:2 or 2:1 InRes [22,27,28]. Different design strategies have been proposed to introduce the quadratic terms needed to promote 1:2 InRes in microbeams, including H- and U-shaped designs with inertial coupling [28], stepped beams [23], and arch beam resonators [29]. Unlike 1:3 InRes, which is characterized by strong hardening, microbeam resonators with 1:2 or 2:1 InRes exhibit its signature M-shaped resonance curves. Refs. [18,23] suggested that this behavior can reduce Allan deviation of frequency to a certain extent when 1:2 InRes is triggered. More recently, Jun et al. demonstrated that frequency saturation can also be achieved through 1:2 InRes if strong Duffing nonlinearity is introduced along with quadratic coupling [22]. A group of studies also examine InRes in structures at macro level [30–32] in the context of structural dynamics. Those demonstrate the qualitative behavior of structures when InRes is triggered, such as isola formations, with no attention given to the frequency stabilization perspective.

Despite these developments, several key questions remain unresolved. Specifically, (i) how do different orders of nonlinearity shape the frequency response functions when either 1:2 or 1:3 InRes is triggered, (ii) what are the critical parameters that determine stabilization performance, and (iii) is a zero-dispersion ($d\omega/dE = 0$) point essential for achieving frequency stabilization? To address these questions, this paper compares 1:2 and 1:3 InRes in systems exhibiting hardening Duffing nonlinearity, with a harmonic drive applied to the lower frequency mode. While closed-loop operation to create self-sustained oscillators is typically used in practical applications for timing and sensing, we focus exclusively on open-loop operation for this analysis. Nonetheless, our study centers on the phase difference of π/2 between the driving force and the first mode, as this configuration closely mirrors the conditions used in closed-loop systems. First, we examine how various system parameters, such as types of InRes and interaction between coupling and hardening nonlinearity, affect frequency response curves (FRCs). We then explore the implications of FRC shapes on frequency stabilization mechanisms and examine the influence of each parameter.

## 2. Theoretical Model and Solution Methodology

We study 1:*n* InRes in a microbeam using a two-degree-of-freedom reduced order dynamic model that incorporates coupling and hardening stiffness terms. Since the actuation and geometric nonlinearities are continuously distributed along the beam, developing accurate reduced-order models with appropriate nonlinear terms is a nontrivial task [33,34]. In this study, we intend to employ the simplest model that effectively captures the dynamic interaction of InRes and the hardening effect, using the fewest but most critical variables. This approach allows for detailed parametric studies to provide deeper insights into the stabilization mechanism.

The equations of motion of the externally driven mode $u_1(t)$ and internally resonant higher mode $u_2(t)$ are given as:

$$\ddot{u}_1 + \omega_1^2 u_1 + 2\zeta_1\omega_1\dot{u}_1 + \gamma_1 u_1^3 + \frac{\partial H_c}{\partial u_1} = f(t), \tag{1a}$$

$$\ddot{u}_2 + \omega_2^2 u_2 + 2\zeta_2\omega_2\dot{u}_2 + \frac{\partial H_c}{\partial u_2} = 0 \tag{1b}$$

where $\gamma_1 > 0$ is the cubic term representing the hardening effect due to midplane stretch of the beam, $H_c$ is the Hamiltonian that governs the interaction between the two modes. In the case of 1:$n$ InRes, i.e., $\omega_1 \approx n\omega_2$, the Hamiltonian can be express as $H_c = \alpha u_1^n u_2$ where $\alpha$ is the coupling constant. Note that the hardening effect is considered only in the externally driven mode ($u_1$), while the second mode ($u_2$) is assumed to be linear when the nonlinear coupling term is excluded. In the open loop operation of a micromechanical resonator, the first mode is externally excited by $f(t) = F \sin(\Omega t)$ at a frequency $\Omega$ close to its first mode freqency. Assuming that the first mode is externally driven by the harmonic force and the second mode is internally driven by the first mode, the steady-state solutions to Eq. (1) can be assumed to take the following harmonic forms:

$$u_1 = U_1 \sin(\psi), \quad u_2 = U_2 \sin(n\psi + \beta), \tag{2a,b}$$

where $\psi = \Omega t + \phi_1$, $\beta$ is the relative phase between the modes, and $U_i$ and $\phi_i$ are the steady-state amplitude and phase of mode $i$, respectively.

Substituting Eq. (2) into Eq. (1) and balancing the corresponding harmonic terms, we obtain the following set of nonlinear algebraic equations for 1:$n$ InRes where $n = 2, 3$:

$$(\omega_1^2 - \Omega^2)U_1 + \frac{3}{4}\gamma_1 U_1^3 - \frac{n\alpha}{2}U_1^{n-1}U_2 \cos(\beta - \theta_n) = F \cos\phi_1, \tag{3a}$$

$$2\zeta_1 \omega_1 \Omega U_1 + \frac{n\alpha}{2}U_1^{n-1}U_2 \sin(\beta - \theta_n) = -F \sin\phi_1. \tag{3b}$$

Here, $\theta_n$ is a coefficient given as $\theta_3 = 0$ for 1:3 IR and $\theta_2 = \pi/2$ for 1:2 IR. As the second mode is assumed to stay within linear regime [15], Eq. (1b) can be rewritten by considering the assumed solution of first mode $u_1$ in Eq. (2a) as the excitation to the linear second mode:

$$\ddot{u}_2 + \omega_2^2 u_2 + 2\zeta_2 \omega_2 \dot{u}_2 = \alpha \frac{(4-n)}{4} U_1^n \sin(n\psi - \theta_n). \tag{4}$$

The steady-state amplitude $U_2$ and the relative phase between vibratory modes $\beta$ can be then found as:

$$U_2 = \frac{(4-n)\alpha U_1^n}{4\sqrt{(2\zeta_2\omega_2(n\Omega))^2 + (\omega_2^2 - (n\Omega)^2)^2}}, \quad \beta = \theta_n + \tan^{-1}\frac{2\zeta_2\omega_2(n\Omega)}{\omega_2^2 - (n\Omega)^2}, \quad n = 2, 3, \tag{5a,b}$$

Combining Eqs. (3a,b) by utilizing $\sin\phi_1^2 + \cos\phi_1^2 = 1$ gives the frequency-amplitude relationship:

$$\left[(\omega_1^2 - \Omega^2)U_1 + \frac{3}{4}\gamma_1 U_1^3 - \frac{n\alpha}{2}U_1^{n-1}U_2\cos(\beta - \theta_n)\right]^2 + \left[2\zeta_1\omega_1\Omega U_1 + \frac{n\alpha}{2}U_1^{n-1}U_2\sin(\beta - \theta_n)\right]^2 - F^2 = 0 \tag{6}$$

Substituting Eq. (5) into Eq. (6) results in a nonlinear algebraic equation in terms of $U_1$ for given forcing amplitude and frequency ($F, \Omega$). This nonlinear algebraic equation is solved by employing the Newton-Raphson solution scheme, equipped with arc-length continuation to effectively capture the coexisting of multiple solutions and turning points of the frequency response function. Further details regarding this solution methodology can be found in Ref. [35]. Once the response amplitude of the first mode $U_1$ is obtained, the second mode response amplitude $U_2$ and corresponding phasing values $\phi_i$ can be determined by back-substituting $U_1$ into Eqs. (3,5).

The presence of multiple solutions and turning points in this nonlinear system indicates bifurcation behavior. In particular, cubic stiffness nonlinearity combined with modal coupling terms is known to cause two types of bifurcations: saddle-node (SN) bifurcations, typically observed in Duffing curves, and Hopf bifurcations (HB). These bifurcation points can be identified by introducing perturbations $v_i$ around the steady-state solution $u_i$ and evaluating the Floquet multipliers using the monodromy matrix [22]. Floquet analysis enables us not only to identify the stability of the periodic solutions but also to predict transitions between different dynamic regimes.

## 3. Discussion on the frequency response

Building on the solutions derived in Section 2, this section presents an in-depth analysis of the FRCs under varying system parameters. By examining how these parameters influence the shapes of the FRCs, we aim to uncover their role in the frequency stabilization mechanism. Emphasis will be placed on the condition where the phase $\phi_1 = \pi/2$ is achieved, which serves as a critical reference for understanding the relationship between amplitude and frequency when the micromechanical resonator is operated in a closed-loop configuration to achieve self-sustained oscillations.

In nonlinear FRCs, the backbone curve – obtained by setting damping and forcing to zero – typically represents the frequency-amplitude relationship without the influence of damping and external forcing. Here, we instead derive the frequency-amplitude relationship under the condition of $\phi_1 = \pi/2$ which we refer to as the $\pi/2$–backbone curve. As the name suggests, this curve is conceptually similar to the traditional backbone curve; however, damping of the second mode ($\zeta_2$) is not neglected. This curve physically represents the condition under which the peak amplitude ($U_{1p}$) and corresponding peak frequency ($\Omega_{1p}$) are achieved in open-loop operation across different forcing levels. It also reflects the conditions under which self-sustained oscillations are maintained in a closed-loop operation where $\phi_1 = \pi/2$ is facilitated in the phase-locked loop. For a linear system without modal coupling and Duffing nonlinearity (i.e., $\alpha = 0$, $\gamma_1 = 0$), this condition corresponds to resonance as $U_1$ is maximized in Eq. 3(b) for $\phi_1 = \pi/2$. In the absence of modal coupling but in the presence of Duffing nonlinearity (i.e., $\alpha = 0$, $\gamma_1 \neq 0$), this curve traces the location where the jump-down phenomena occur as the forcing level varies, making it technically analogous to the traditional backbone curve.

By setting $\phi_1 = \pi/2$ in Eq. (3a), the equations of the $\pi/2$–backbone curve is obtained as below for 1:2 InRes case:

$$\Omega_p^2 = \omega_1^2 + U_{1p}^2 \left[ \frac{3}{4}\gamma_1 + \frac{\alpha^2}{2} G_2[\Omega_p] \right] \text{ where } G_2[\Omega_p] = \frac{(4\Omega_p^2 - \omega_2^2)}{(4\zeta_2\omega_2\Omega_p)^2 + (4\Omega_p^2 - \omega_2^2)^2}. \quad (7)$$

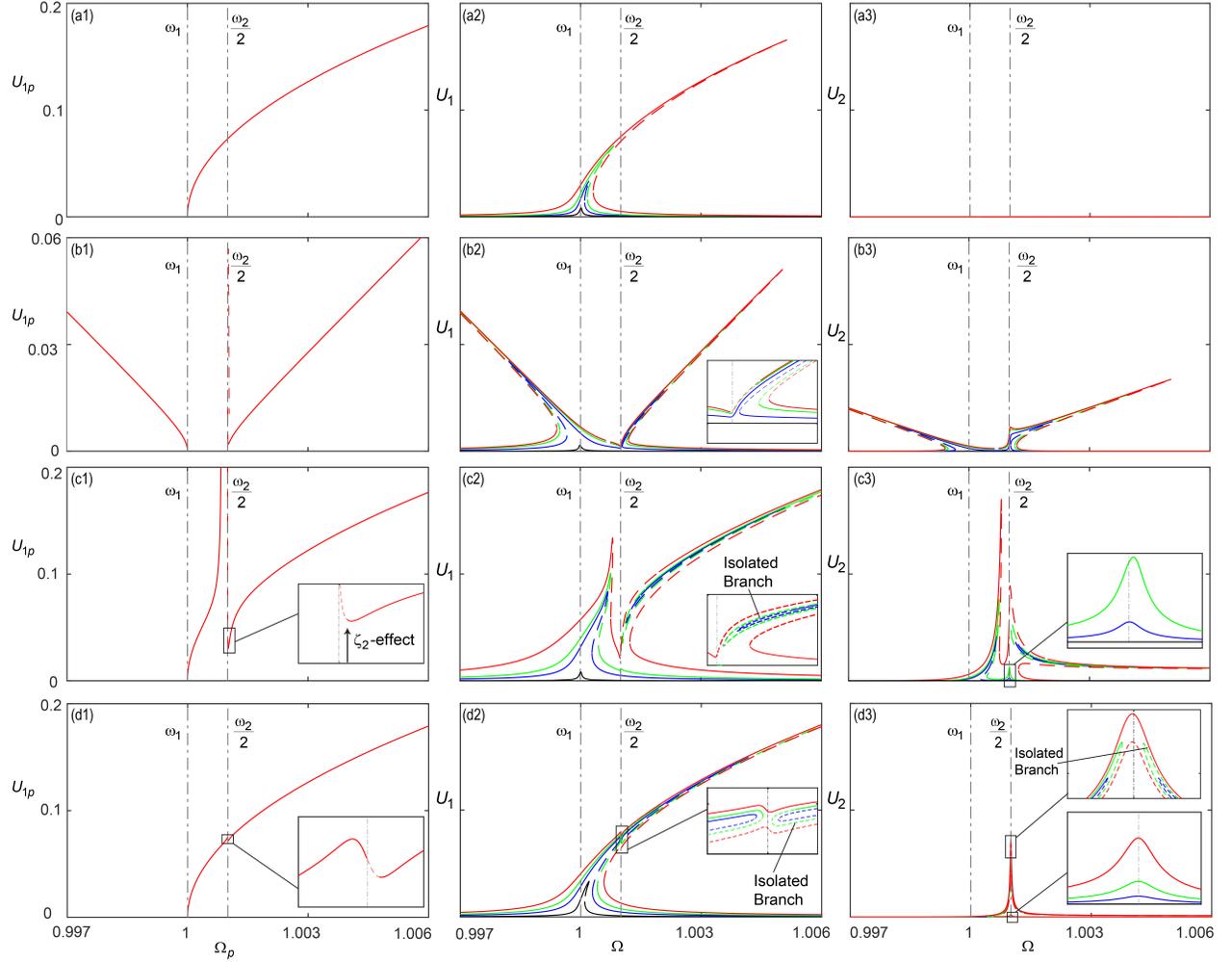

**Fig. 1** Comparison of π/2–backbone curve and frequency response curves for 1:2 InRes under different conditions: (a) a Duffing oscillator with $\gamma_1 = 0.5$, $\alpha = 0$, (b) a system with only intermodal coupling with $\gamma_1 = 0$, $\alpha = 0.5$, (c) a system that includes Duffing nonlinearity and strong coupling $\gamma_1 = 0.5$, $\alpha = 0.03$, and (d) a system that includes cubic and weak coupling $\gamma_1 = 0.5$, $\alpha = 0.03$. Here, the mode frequencies and damping are set as $\omega_1 = 1$, $\omega_2 = 2.002$, and $\zeta_1\omega_1 = \zeta_2\omega_2 = 3 \times 10^{-5}$. The first column (a1-d1) shows π/2–backbone curves, the second column (a2-d2) shows the forced response of the first mode $U_1$, and the third column (a3-d3) shows the second mode response ($U_2$), at various forcing levels $F$. Dashed lines represent unstable solutions.

Eq. (7) can be rearranged to provide the amplitude $U_{1p}$ at the corresponding frequency $\Omega_p$:

$$U_{1p}^2 = \frac{\Omega_p^2 - \omega_1^2}{\frac{3}{4}\gamma_1 + \frac{\alpha^2}{2}G_2[\Omega_p]} \tag{8}$$

Figure 1 compares the $\pi/2$–backbone curves of 1:2 InRes obtained for four different cases: (i) a Duffing oscillator with $\gamma_1 = 0.5$, $\alpha = 0$, (ii) a system with only intermodal coupling with $\gamma_1 = 0$, $\alpha = 0.5$, and (iii-iv) a system that includes both cubic term of $\gamma_1 = 0.5$ and intermodal coupling terms $\alpha = 0.03$ and $\alpha = 0.003$, respectively. Here, the mode frequencies are set to be $\omega_2 > 2\omega_1$ ($\omega_1 = 1$ and $\omega_2 = 2.002$) to have the effect of InRes manifest in the hardening region.

In the absence of coupling term ($\alpha = 0$), Eq. (7) reduces to the backbone curve equation for a typical hardening behavior with a strong amplitude-frequency effect:

$$\Omega_p^2 = \omega_1^2 + \frac{3}{4}U_{1p}^2\gamma_1 \tag{9}$$

as shown in Fig. 1(a). For the case where only intermodal coupling is considered ($\gamma_1 = 0$), the $\pi/2$–backbone curve equation is given by

$$\Omega_p^2 = \omega_1^2 + \frac{U_{1p}^2\alpha^2}{2}\frac{4\Omega_p^2 - \omega_2^2}{(4\zeta_2\omega_2\Omega_p)^2 + (4\Omega_p^2 - \omega_2^2)^2} \quad \text{or} \tag{10a}$$

$$U_{1p}^2 = \left(\Omega_p^2 - \omega_1^2\right) \Big/ \frac{\alpha^2}{2}G_2[\Omega_p] \tag{10b}$$

Thus, Fig. 1(b) exhibits two solution branches: one with softening when $\Omega_p < \omega_2/2$ and another with hardening when $\Omega_p > \omega_2/2$. Note that no solution exists in the range of $\omega_1 < \Omega_p < \omega_2/2$ as $G_2[\Omega_p] < 0$ in Eq. 12(b).

Figure 1(c) shows the $\pi/2$–backbone when both cubic and coupling terms are present ($\gamma_1 \neq 0$, $\alpha \neq 0$). The softening branch of the 1:2 InRes shown in Fig. 1(b) is observed to bend to the right in Fig. 1(c) due to the inclusion of the cubic hardening term. This balance between the InRes softening and the Duffing hardening effects eliminates the amplitude-frequency effect within the narrow frequency range near $\omega_2/2$. This ensures that variations in amplitude $U_{1p}$ do not shift the frequency away from $\omega_2/2$. When $\Omega_p < \omega_2/2$, the curve leads to a vertical asymptote ($\Omega_{asy}$), arising from a singularity in Eq. (8). For such cases, the vertical asymptote can be approximated by assuming $\zeta_2 = 0$:

$$\Omega_{asy}^2 = \left(\frac{\omega_2}{2}\right)^2 - \frac{\alpha^2}{6\gamma_1} \tag{11}$$

$\Omega_{asy}$ is close to $\omega_2/2$ when the term $\alpha^2/6\gamma_1$ is small, but there exists a gap of $\alpha^2/6\gamma_1$ from $\omega_2/2$ when the value is not negligible. When $\Omega_p > \omega_2/2$, the $\pi/2$–backbone curve converges to the Duffing curve as the effect of coupling term, $G_2[\Omega_p]$, diminishes according to Eq. (10a). When $\zeta_2 \approx 0$, we also observe that the amplitude converge to zero at $\Omega_p = \omega_2/2$, facilitating the formation of the asymptote. It is important to note that damping can pull this branch up, as shown in the inset of Fig. 1(c3).

If the damping is sufficiently strong, the asymptote can even disappear. The following criterion can be obtained to ensure the formation of the asymptote. Rewriting Eq.(10b) yields

$$U_{1p} = \sqrt{\frac{\Omega_p^2 - \omega_1^2}{\left[\frac{3}{4}\gamma_1 + \frac{\alpha^2}{2}G_2[\Omega_p]\right]}} \tag{12}$$

where the function $G_2$ reaches its minimum $\left(\frac{dG_2}{d\Omega_p}\Big|_{\Omega_p^*} = 0\right)$ at $\Omega_p^* = \omega_2\sqrt{1-2\zeta_2}/2$ such that:

$$\min\{G_2[\Omega_p]\} = \frac{-1}{4\zeta_2\omega_2^2(1-\zeta_2)} \tag{13}$$

Considering the singularity of Eq.12 at $\Omega_p^* = \omega_2\sqrt{1-2\zeta_2}/2$, the following criterion can be defined to ensure that the coupling strength relative to the Duffing constant ($\alpha^2/\gamma_1$) is large enough to overcome the second mode damping effect to form an asymptote:

$$\frac{\alpha^2}{\gamma_1} > 6\omega_2^2\zeta_2(1-\zeta_2) \tag{14}$$

Figure 1(d) demonstrates a case where the coupling is not strong enough to overcome the damping effect of the second mode, thereby eliminating the vertical asymptote $\Omega_{asy}$. In this case, the $\pi/2$–backbone curve does not differ significantly from the Duffing curve, except for slight deviations near the InRes frequency $\omega_2/2$. These deviations manifest as localized undulations in the curve, as shown in the inset of Fig. 1(d), indicating the influence of the coupling effect around $\Omega \approx \omega_2/2$. Based solely on the $\pi/2$–backbone curve, it is difficult to anticipate the frequency stabilization effect. However, further analysis of the FRCs and bifurcation reveals that frequency and amplitude saturation phenomena occur around $\omega_2/2$ over a range of forcing amplitudes, forming the basis for the frequency stabilization, even with such a very small coupling effect, as detailed in the subsequent discussion.

The second and third columns in Fig. 1 show the FRCs of both the first and second modes at various forcing amplitudes for 1:2 InRes in each case. Figures 1(a-b) follow the well-documented FRCs for Duffing oscillators and 1:2 InRes. Figure 1(c) represents the FRCs with InRes and Duffing nonlinearity ($\gamma_1 = 0.5$, $\alpha = 0.03$, $\zeta_1\omega_1 = \zeta_2\omega_2 = 3\times10^{-5}$), where the asymptote exists according to Eq.(14). For the sake of clarity, refer to Fig. 3(d1), which displays the same FRCs at an adjusted scale. At lower forcing amplitudes, the resonance is formed away from $\omega_2/2$ due to frequency mismatch (i.e., $\omega_1 = 1$, $\omega_2 = 2.002$) and exhibits typical hardening with two SN bifurcations and a strong amplitude-frequency effect. The response of the second mode in this range is not noticeable, indicating weak modal interactions between vibratory modes. When the forcing amplitude increases sufficiently to bring the FRC into the 1:2 InRes zone, as seen in the blue and green curves, the FRC splits into two parts near $\Omega \approx \omega_2/2$, with an isolated branch forming at $\Omega > \omega_2/2$. This isolated solution branch, also known as an isola [36], would not be observed in frequency sweep-up experiments, as the forced response jumps down to the lower branch near $\Omega_{asy}$. Also observed here is an increase in the second mode amplitude for both upper and isolated branches. While only the first mode is externally excited, the second mode is apparently excited internally through intermodal coupling, drawing energy from the first mode. A further increase in the forcing amplitude merges the isola with the lower branch near $\Omega \approx \omega_2/2$, as seen in the red curve. In this scenario, closed-loop operation with a $\pi/2$ phase shift would follow the $\pi/2$–backbone, shown in Fig. 1(c1), where the frequency asymptotically approaches and eventually saturates at $\Omega_{asy}$ while the amplitude continues to increase.

When the coupling effect is weak, as shown in Fig. 1(d), even though the asymptote line disappears in the $\pi/2$–backbone curve, the intermodal coupling still separates the FRC near $\Omega \approx \omega_2/2$, yielding an isolated upper branch. For clarity, refer to Fig. 3(c1), which shows the same FRCs at an adjusted scale. Once the seperation of the FRC occurs, the forced response jumps down to the lower branch at the separation point near $\Omega \approx \omega_2/2$ during the frequency sweep-up operation. Then, the isolated branch existing at $\Omega > \omega_2/2$ does not appear. Further increseas in the forcing amplitude do not significantly alter the response amplitude and jump-down frequency, leading to frequency and amplitude saturation phenomena over a range of forcing amplitudes. Thus, even with very small coupling, InRes serves as a robust mechanism for stabilizing frequency through the saturation of both frequency and amplitude. When the forcing amplitude is further increased beyond a critical point ($1.8 \times 10^{-5}$), as shown in the red curve, the isola merges with the main curve and exhibits HB after passing thorough $\Omega \approx \omega_2/2$. After this merge, the FRC becomes similar to typical Duffing curves. Note that the experimental results in Ref [22] demonstrated the frequency and amplitude saturation phenomena described here.

We now apply a similar approach to analyze the 1:3 InRes. First, the equation of the $\pi/2$–backbone curve is obtained by imposing $\phi_1 = \pi/2$ in Eq.3(a), leading to a second-order polynomial in terms of $U_{1p}^2$:

$$U_{1p}^4 \left( \frac{3\alpha^2}{16} G_3[\Omega_p] \right) + U_{1p}^2 \left( \frac{3}{4}\gamma_1 \right) + \left( \omega_1^2 - \Omega_p^2 \right) = 0 \tag{15a}$$

$$\text{where} \quad G_3[\Omega_p] = \frac{(9\Omega_p^2 - \omega_2^2)}{(6\zeta_2\omega_2\Omega_p)^2 + (9\Omega_p^2 - \omega_2^2)^2} \tag{15b}$$

Thus, the $\pi/2$–backbone for 1:3 InRes allows coexisting of two solutions of $U_{1p}^2$ at a given $\Omega_p$. The solution of Eq. (15) can be explicitly obtained as:

$$U_{1p}^2 = \frac{-2\gamma_1}{\alpha^2 G_3[\Omega_p]} \pm \sqrt{\left( \frac{2\gamma_1}{\alpha^2 G_3[\Omega_p]} \right)^2 + \left( \frac{16}{3\alpha^2 G_3[\Omega_p]} \right)(\Omega_p^2 - \omega_1^2)} \tag{16}$$

Further simplification into the backbone curve can be performed by setting $\zeta_2 = 0$:

$$U_{1p}^2 = -\frac{2\gamma_1\sigma}{\alpha^2} \pm \sqrt{\left( \frac{2\gamma_1\sigma}{\alpha^2} \right)^2 + \frac{16\sigma}{3\alpha^2}(\Omega_p^2 - \omega_1^2)} \quad \text{where} \quad \sigma = (3\Omega_p)^2 - \omega_2^2 \tag{17}$$

Similar to the 1:2 InRes case, Fig. 2 compares $\pi/2$–backbone curves of 1:3 InRes, obtained for four different cases at $\omega_1 = 1$, $\omega_2 = 3.015$: (i) a Duffing oscillator with $\gamma_1 = 0.5$, $\alpha = 0$, which is exactly the same as Fig. 1a, (ii) a system with only intermodal coupling with $\gamma_1 = 0$, $\alpha = 20$, and (iii-iv) a system that includes both Duffing $\gamma_1 = 0.5$ and intermodal coupling terms $\alpha = 0.1$ and $\alpha = 0.002$, respectively. In the absence of Duffing nonlinearity ($\gamma_1 = 0$), the backbone curve is given by:

$$\Omega_p^2 = \omega_1^2 + U_{1p}^4 \frac{3\alpha^2}{16} \frac{\sigma}{(6\zeta_2\omega_2\Omega_p)^2 + \sigma^2} \tag{18}$$

Comparing the pure 1:2 InRes (Fig. 1b) and 1:3 InRes (Fig. 2b), the cubic coupling in 1:3 InRes provides higher-order curve shapes, as indicated by comparing Eq. (10a) and Eq. (18), in both softening and hardening backbone curves for $\Omega_p < \omega_1$ and $\Omega_p > \omega_2/3$, respectively.

Figure 2(c) shows the 1:3 InRes $\pi/2$–backbone curves when both Duffing and coupling terms are included ($\gamma_1 = 0.1$, $\alpha = 0.1$). As predicted by Eq. (19), when $\Omega_p > \omega_2/3 > \omega_1$ or $\Omega_p < \omega_1 < \omega_2/3$, there exists only one positive solution for $U_{1p}^2$, while there can exist two positive solutions in the range of $\omega_1 < \Omega_p < \omega_2/3$. Also, the term inside the square root becomes negative and no solution exists in the range of $\kappa\omega_1 < \Omega_p < \omega_2/3$ where $\kappa = \sqrt{\dfrac{4\alpha^2\omega_1^2 + 3\gamma_1^2\omega_2^2}{4\alpha^2 + 27\gamma_1^2}}$. Considering these findings, the $\pi/2$–backbone curve in Fig. 2c displays a mixed hardening and softening behavior with the coexistence of multiple solutions for $U_{1p}$. Unlike the asymptotic approach of the 1:2 InRes backbone to $\Omega_{\text{asy}}$, there exists a point where the slope of $\dfrac{\partial \Omega_p}{\partial U_p}$ becomes zero at $\Omega_p = \kappa\omega_1$, which is referred to as a zero-dispersion point in Ref. [6,15].

The FRCs for this case are shown in the second and third columns in Fig. 2(c). The intermodal coupling once again splits the FRC into two parts near $\Omega_p \approx \omega_2/3$, forming an isola at $\Omega_p > \omega_2/3$. One noticeable difference in the range of $\Omega_p < \omega_2/3$ is the existence of three SN bifurcations due to the coexistence of multiple backbone solutions (refer to Fig. 4(d1) for enhanced clarity). During the open-loop sweep-up operation, the forced response is expected first to exhibit a jump-up, followed by a jump-down phenomenon. In closed-loop operation, however, the self-sustained oscillations are maintained along the $\pi/2$–backbone curve. In this case, while the amplitude-frequency effect is reduced at the zero-dispersion point, larger oscillations beyond this point still suffer from this effect, governed by the shape of the $\pi/2$– backbone curve. As will be discussed later, the coupling constant ($\alpha$) significantly influences this shape and, consequently, the effectiveness of the frequency stabilization.

Similar to the 1:2 InRes case, considering the maximum of the function $G_3$ given in Eq.(18), the criterion for the branch separation can be expressed as:

$$\frac{\alpha^2}{\gamma_1^2} > \frac{27\omega_2^2 \zeta_2(1-\zeta_2)}{\omega_2^2(1-2\zeta_2) - 9\omega_1^2} \tag{19}$$

When the coupling constant relative to the Duffing constant (i.e., $\alpha^2/\gamma_1^2$) is not strong enough, the $\pi/2$–backbone curve remains continuous, without separation and zero-dispersion, resulting in qualitatively similar $\pi/2$–backbone curves and FRCs for both 1:2 and 1:3 InRes cases, as shown in Fig. 1(d) and Fig. 2(d). Under this condition, we again observe the frequency and amplitude saturation phenomenon together, which is effective for stabilizing frequencies. After reviewing previous experimental studies that demonstrated frequency stabilization via InRes [18,22,37], it appears that their experimental conditions align with this case. This is because both amplitude and frequency saturation phenomena were observed when InRes occurred, similar to what is seen in this weak coupling scenario.

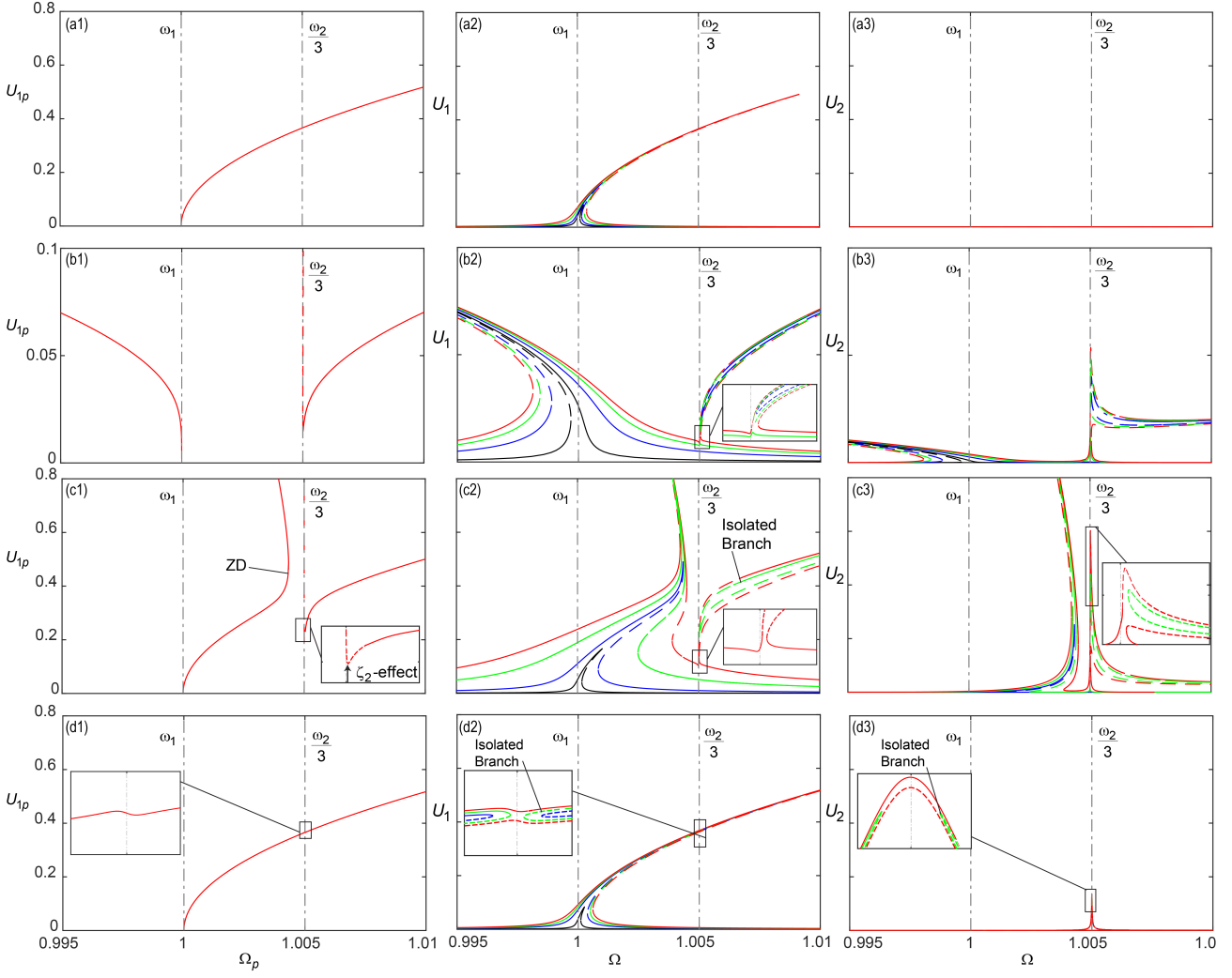

**Fig. 2** Comparison of π/2–backbone curve and frequency response curves for 1:3 InRes under different conditions: (a) a Duffing oscillator with $\gamma_1 = 0.1$, $\alpha = 0$, (b) a system with intermodal coupling only with $\gamma_1 = 0$, $\alpha = 20$, (c) a system that includes Duffing nonlinearity and strong coupling $\gamma_1 = 0.1$, $\alpha = 0.1$, and (d) a system that includes Duffing nonlinearity and weak coupling $\gamma_1 = 0.1$, $\alpha = 0.002$. Here, the mode frequencies and damping are set as $\omega_1 = 1$, $\omega_2 = 3.015$ and $\zeta_1\omega_1 = \zeta_2\omega_2 = 3\times 10^{-5}$. The first column (a1-d1) shows π/2–backbone curves, the second column (a2-d2) shows the forced response of the first mode $U_1$, and the third column (a3-d3) shows the second mode response $U_2$, at various forcing levels $F$. Dashed lines represent unstable solutions.

## 4. Parametric Studies on Frequency Stabilization

This section presents a parametric study to examine the effect of system parameters on the steady-state response in the context of frequency stabilization. We systematically investigate how these parameters influence the peak frequency ($\Omega_p$) when the system operates at the phase of $\pi/2$. Here, we assume that frequency fluctuations are primarily caused by variations in the peak amplitude of first mode response ($U_{1p}$) and the forcing amplitude ($F$). Thus, analyzing the sensitivity of the peak frequency $\Omega_{1p}$ to both the peak amplitude $U_{1p}$ and the forcing amplitude $F$ is crucial. While FRCs and $\pi/2$–backbone curves are useful for examining sensitivity to amplitude $U_{1p}$, they do not explicitly capture the influence of the forcing amplitude $F$. To address this limitation, we also present the results plotting the relationship between the peak frequency $\Omega_{1p}$ and the forcing amplitude $F$. These graphs provide a clearer illustration of frequency stability under varying forcing amplitudes and help elucidate the underlying stabilization mechanisms. In section 3, we identified different patterns of frequency stabilization depending on the coupling strength, which can be classified into two distinct regimes: (i) weak coupling and (ii) strong coupling. We begin by summarizing the frequency stabilization mechanisms under these two regimes in greater detail for both 1:2 and 1:3 InRes, and further investigate the influence of coupling constant and frequency mismatch.

Figure 3 shows the results for 1:2 InRes with weak ($\alpha = 0.003$) and strong ($\alpha = 0.03$) coupling at $\omega_1 = 1$, $\omega_2 = 2.002$, $\gamma_1 = 0.5$, and $\omega_1 \zeta_1 = \omega_2 \zeta_2 = 3 \times 10^{-5}$. In Fig. 3(a), the $\pi/2$–backbone curves are compared for the weak and strong coupling cases. As previously discussed, the weak coupling case does not significantly deviate from the Duffing curve, showing a strong amplitude-frequency effect in its $\pi/2$–backbone curve. However, the corresponding FRCs in Fig. 3(c1) reveal a separation of the curve into two parts near $\Omega \approx \omega_2/2$, leading to saturation in the jump-down frequency as the forcing amplitude $F$ increases.

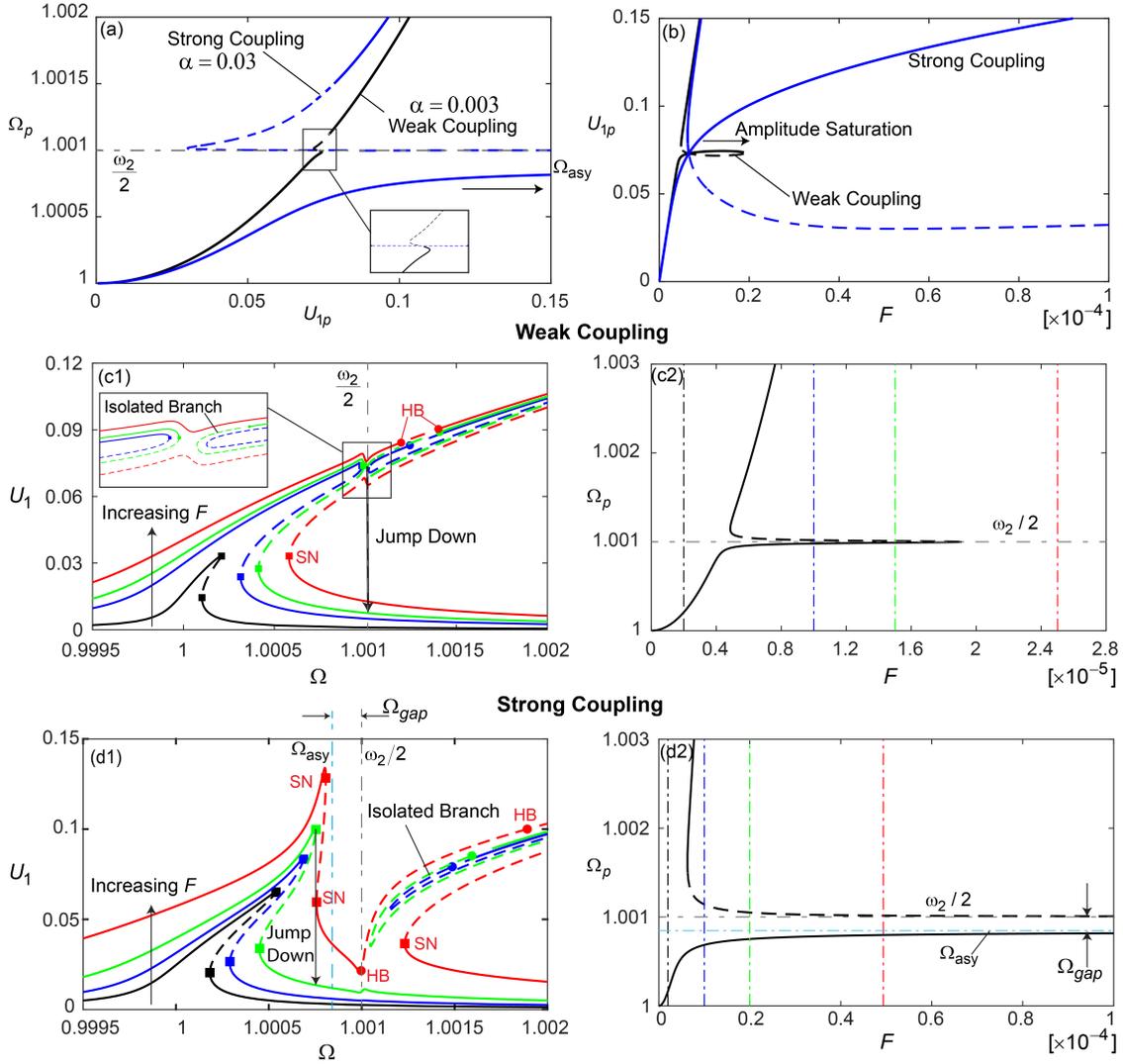

**Fig. 3** Comparison of dynamic behaviors in the regime of weak ($\alpha = 0.003$) and strong ($\alpha = 0.03$) coupling for 1:2 InRes at $\omega_1 = 1$, $\omega_2 = 2.002$, $\gamma_1 = 0.5$, and $\omega_1\zeta_1 = \omega_2\zeta_2 = 3\times 10^{-5}$. (a) shows the $\pi/2$–backbone curve for weak and strong coupling, illustrating how the peak frequency $U_{1p} = 0.07$, $\Omega_p = \omega_2/2$ varies with the peak amplitude $U_{1p}$. (b) plots the peak amplitude $U_{1p}$ as a function of forcing amplitude $F$, highlighting amplitude saturation for the weak coupling case only. For the weak coupling regime, (c1) shows the frequency response curves (FRCs) at various forcing amplitudes $F$, while (c2) shows the peak frequency $\Omega_p$ versus forcing amplitude $F$. The dashed-dotted lines in (c2) indicate the forcing amplitudes corresponding to the FRCs of the same colors in (c1). Similarly, for the strong coupling regime, (d1) shows the FRCs at various forcing amplitudes, and (d2) plots the peak frequency $\Omega_p$ as a function of forcing amplitude $F$. The dashed-dotted lines in (d2) correspond to the forcing amplitudes used in (d1). Dashed lines indicate unstable solutions. Symbols: (■) Saddle-Node (SN) bifurcations, (●) Hopf bifurcations (HB).

This behavior is further illustrated in the $\Omega_p$ vs. $F$ graph in Fig. 3(c2) for $F \subset [0.5, 1.9] \times 10^{-5}$. Within the saturation region, $\Omega_p$ and $U_{1p}$ are confined to a narrow range by transferring the excessive input energy to the higher mode, resulting in $\partial \Omega_p / \partial F \approx 0$ and $\partial U_{1p} / \partial F \approx 0$ as shown in Figs. 3(a), 3(b) and 3(c2). These observations suggest that stabilization is achieved through *amplitude and frequency saturation*, both pinned around $U_{1p} = 0.07$ and $\Omega_p = \omega_2 / 2$. A further increase in forcing ($F > 1.9 \times 10^{-5}$) causes the isola to merge with the main branch. Beyond this point, frequency and amplitude saturation are lost, and the FRC becomes indistinguishable from the Duffing curve.

The frequency stabilization effect in the strongly coupled case is directly observed from the flat region of the π/2–backbone curve, as shown in Fig. 3(a). In this case, the softening from the InRes coupling is strong enough to counterbalance the Duffing hardening effect, causing the π/2–backbone to asymptotically approach a fixed frequency ($\Omega_{\text{asy}}$). This indicates that frequency becomes less sensitive to fluctuations of $U_{1p}$. The $\Omega_p$ vs. $F$ curve in Fig. 3(d2) shows that the stability region, where $\partial \Omega_p / \partial F \approx 0$ is not constrained by the forcing level, unlike the weak coupling case. It is because the isolated branch merges with the lower solution branch at $\Omega > \omega_2 / 2$. Instead, the frequency saturates to $\Omega_{\text{asy}}$, introducing a frequency gap ($\Omega_{\text{gap}}$) from the InRes frequency of $\omega_2 / 2$. Also, the amplitude does not saturated $\partial U_{1p} / \partial F \neq 0$, but continues to increase with the forcing amplitude, as shown in Figs. 3(b) and 3(d1).

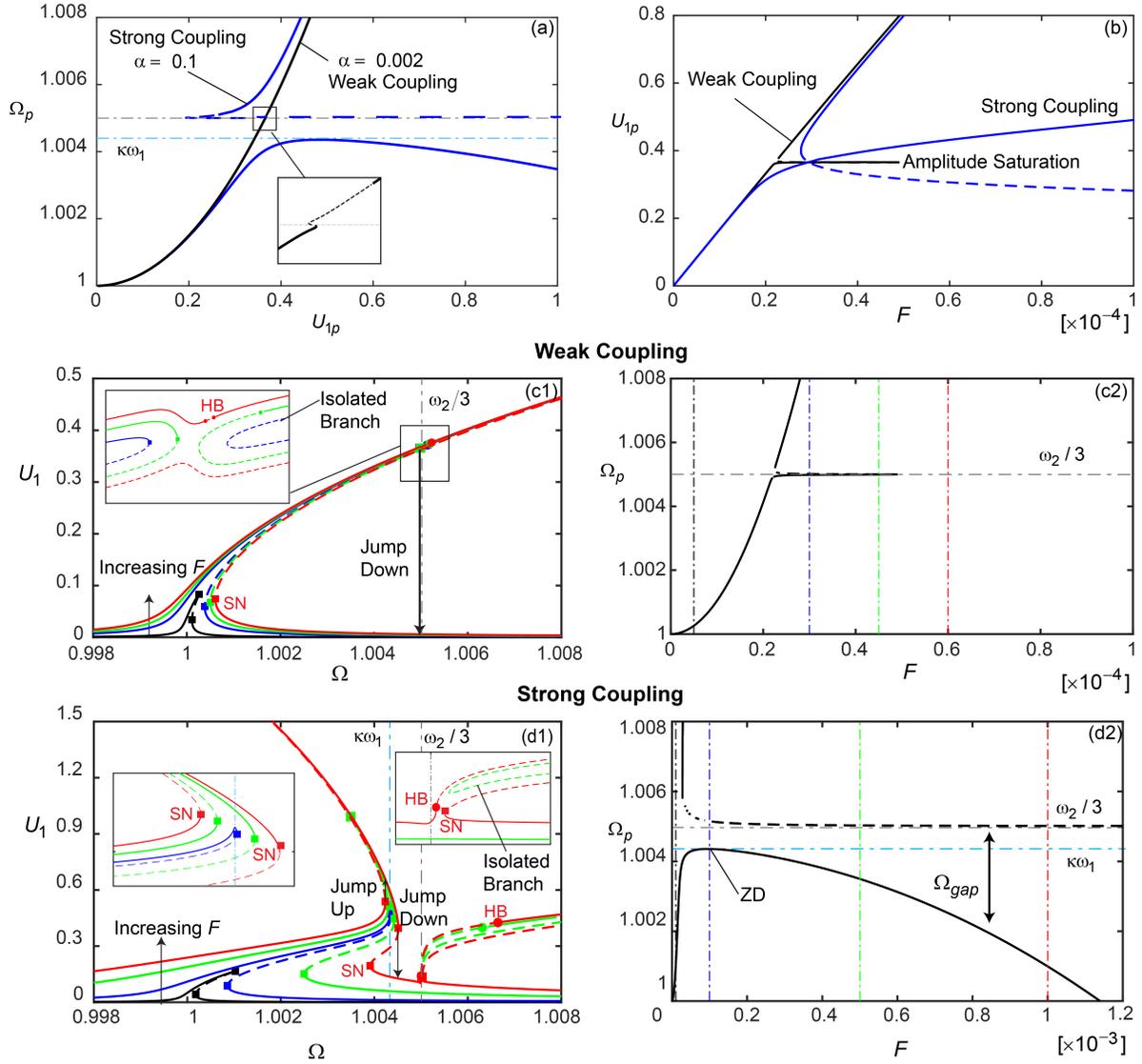

**Fig. 4** Comparison of dynamic behaviors in the regime of weak ( $\alpha = 0.002$ ) and strong ( $\alpha = 0.1$ ) coupling for 1:3 InRes at $\omega_1 = 1$, $\omega_2 = 3.015$ and $\gamma_1 = 0.1$. (a) shows the $\pi/2$–backbone curve for weak and strong coupling, illustrating how the peak frequency $\Omega_p$ varies with the peak amplitude $U_{1p}$. (b) plots the peak amplitude $U_{1p}$ as a function of forcing amplitude $F$, highlighting amplitude saturation for the weak coupling case only. For the weak coupling regime, (c1) shows the frequency response curves (FRCs) at various forcing amplitudes $F$, while (c2) shows the peak frequency $\Omega_p$ versus forcing amplitude $F$. The dashed-dotted lines in (c2) indicate the forcing amplitudes corresponding to the FRCs of the same colors in (c1). Similarly, for the strong coupling regime, (d1) shows the FRCs at various forcing amplitudes, and (d2) plots the peak frequency $\Omega_p$ as a function of forcing amplitude $F$. The dashed-dotted lines in (d2) correspond to the forcing amplitudes used in (d1). Dashed lines indicate unstable solutions. Symbols: (■) Saddle-Node (SN) bifurcations, (●) Hopf bifurcations (HB)

Figure 4 presents the results for 1:3 InRes with weak ($\alpha = 0.002$) and strong ($\alpha = 0.1$) coupling at $\omega_1 = 1, \omega_2 = 3.015$, $\gamma_1 = 0.1$, and $\omega_1 \zeta_1 = \omega_2 \zeta_2 = 3 \times 10^{-5}$. For the regime of weak coupling, the mechanism for frequency stabilization is very similar to the 1:2 InRes case: while the 1:3 InRes does not significantly alter the $\pi/2$–backbone curve from the Duffing curve (see Fig. 4(a)), the separation in the FRC results in the saturation of both amplitude and frequency (i.e., $\partial U_{1p}/\partial F \approx 0$ and $\partial \Omega_p/\partial F \approx 0$) near $\Omega = \omega_2/3$ over a range of range of forcing amplitudes, as shown in Figs. 4(b) and 4(c2). For strong coupling, however, the balance between the 1:3 InRes and the Duffing effect leads to a the zero-dispersion point, $\partial \Omega_p/\partial U_{1p} \approx 0$, in the $\pi/2$–backbone curve. While, for 1:2 InRes, the frequency asymptotically approaches to $\Omega_{asy}$ as forcing increases, the 1:3 InRes results in a bending of the $\Omega_p$ vs. $F$ curve to lower frequencies after passing through the zero-dispersion point. This indicates that frequency stabilization is highly effective around the zero-dispersion, but only over a limited forcing range of forcing around $F \approx 6 \times 10^{-4}$. Based on the provided framework, the frequency saturation behavior experimentally demonstrated in Refs. [18,37] appears to occur in the weakly coupled regime. Thus, the observed frequency stability seems to be primarily influenced by amplitude and frequency saturation effects, rather than by the zero-dispersion effect.

Figure 5 illustrates how coupling strength alters the sensitivity of frequency to the amplitude and forcing variation by varying coupling term $\alpha = \{0, 0.01, 0.1, 0.2\}$ for both 1:2 and 1:3 InRes. When $\alpha = 0.01$, $\pi/2$–backbone curve is still continuous as the criterion given in Eqs. (16) and (21) are not satisfied. While this suggests a strong amplitude-frequency effect for weakly coupled modes, the sensitivity of $\Omega_p$ and $U_{1p}$ to forcing amplitude $F$ is eliminated as evident from the flat region in Figs. 5(a2) and 5(b2). Increasing the coupling term $\alpha$ to 0.1 results in an asymptote and a zero-dispersion for 1:2 and 1:3 InRes, respectively.

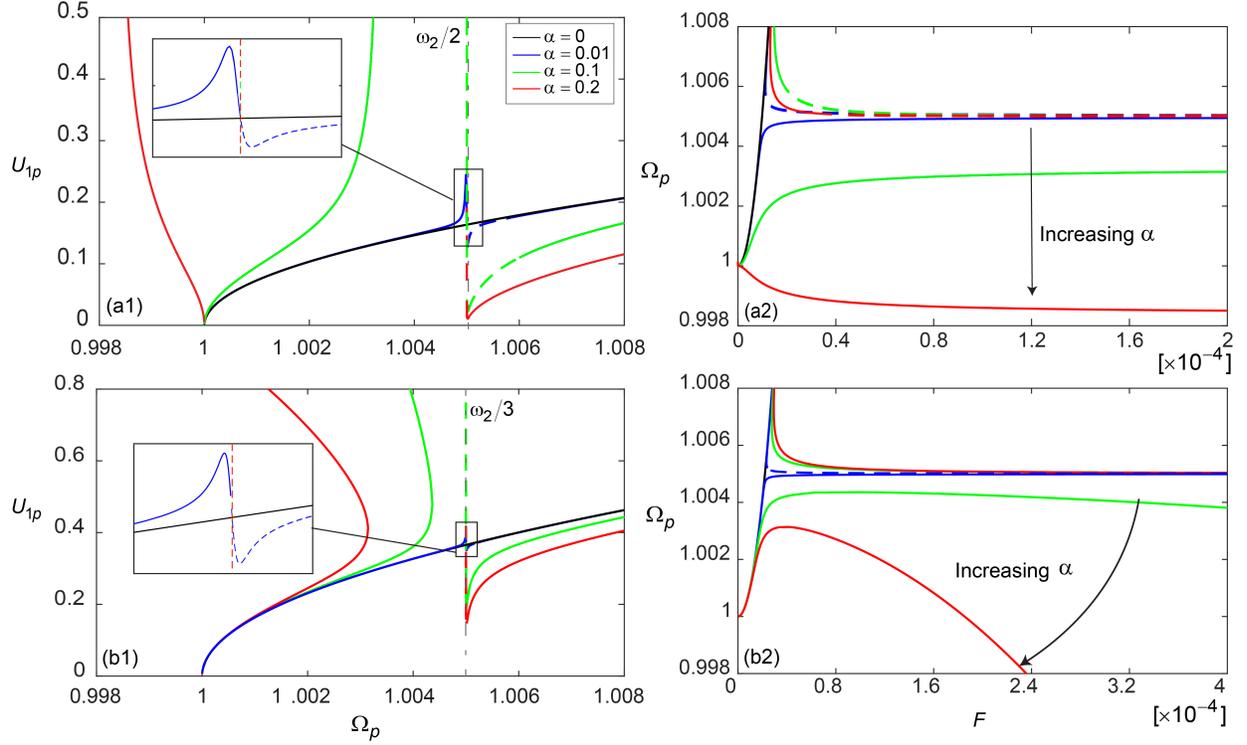

**Fig. 5** Influence of coupling strength on the $\pi/2$–backbone curves for (a) 1:2 InRes at $\gamma_1 = 0.1$, $\omega_1 = 1, \omega_2 = 2.01$ and (b) 1:3 InRes at $\gamma_1 = 0.5, \omega_1 = 1, \omega_2 = 3.015$. (a1,b1) show the peak frequency $\Omega_p$ as a function of peak amplitude $U_{1p}$, and (a2,b2) show the peak frequency $\Omega_p$ as a function of forcing amplitude $F$ for various coupling coefficients $\alpha = \{0, 0.01, 0.1, 0.2\}$. Dashed lines indicate unstable solutions.

Also observed in Fig. 5(a2, b2) is the qualitative difference between the strong coupling of 1:2 and 1:3 InRes. While increasing $\alpha$ merely shifts the stabilization region downward while maintaining its flat shape ($\partial \Omega_p / \partial F \approx 0$) in Fig. 5(a2), the stabilization range of 1:3 InRes is adversely affected by strong coupling. As $\alpha$ increases, it bends the flat shape due to higher-order terms. This suggests there is a range of coupling strength $\alpha$ that provides stabilization for 1:3 InRes, whereas frequency stabilization is achieved at both strong and weak cases for 1:2 InRes.

In Figs. 6 and 7, the effect of frequency mismatch on the stabilization range is explored for both 1:2 and 1:3 InRes. For the negative mismatch case of $\omega_2 = 1.995$ for 1:2 InRes and $\omega_2 = 2.995$ for 1:3 InRes, the InRes does not alter the FRC from the Duffing curve due to its low amplitude. Under perfect matching conditions of 1:2.000 and 1:3.000, signs of modal interactions are observed to some extent, as indicated by the localized undulations in the curve and the increased second mode amplitudes around $\Omega \approx 1$. This interaction, however, is not strong enough to form an isolated branch. This suggests that a positive mismatch is required to achieve stabilization via InRes when $\gamma_1 > 0$ due to the hardening effect. As the mismatch increases, the FRCs begin to separate near $\Omega \approx \omega_2 / n$, and the corresponding $\Omega_{1p}$ vs. $F$ graphs exhibit flat regions, indicating stabilization. Notably, the range of forcing amplitude $F$ over which stabilization occurs expands with increasing mismatch, as experimentally demonstrated in Ref. [22]. This broadening is attributed to the elevation of the energy level at which InRes is activated, which intensifies with the degree of mismatch.

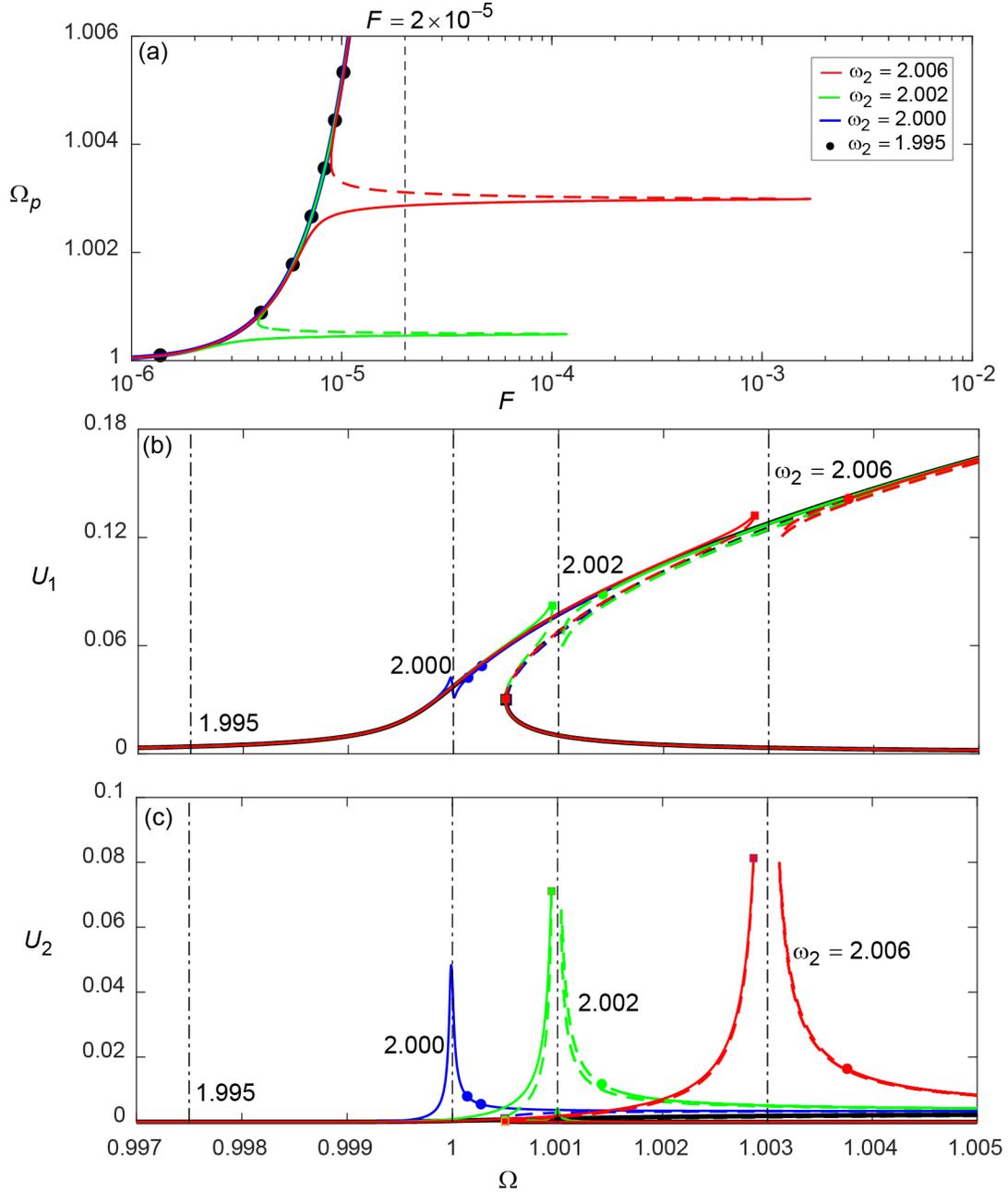

**Fig. 6** Influence of frequency mismatch on the forced response for 1:2 InRes at $\gamma_1 = 0.5, \alpha = 0.1$. (a) shows the peak frequency $\Omega_p$ versus forcing amplitude $F$ for $\omega_2 = [1.995, 2.000, 2.002, 2.006]$. (b) and (c) show the corresponding forced response curves of the first ($U_1$) and second modes ($U_2$), respectively, at $F = 2 \times 10^{-5}$. Dashed lines indicate unstable solutions. Symbols: (■) Saddle-Node (SN) bifurcations, and (●) Hopf bifurcations (HB).

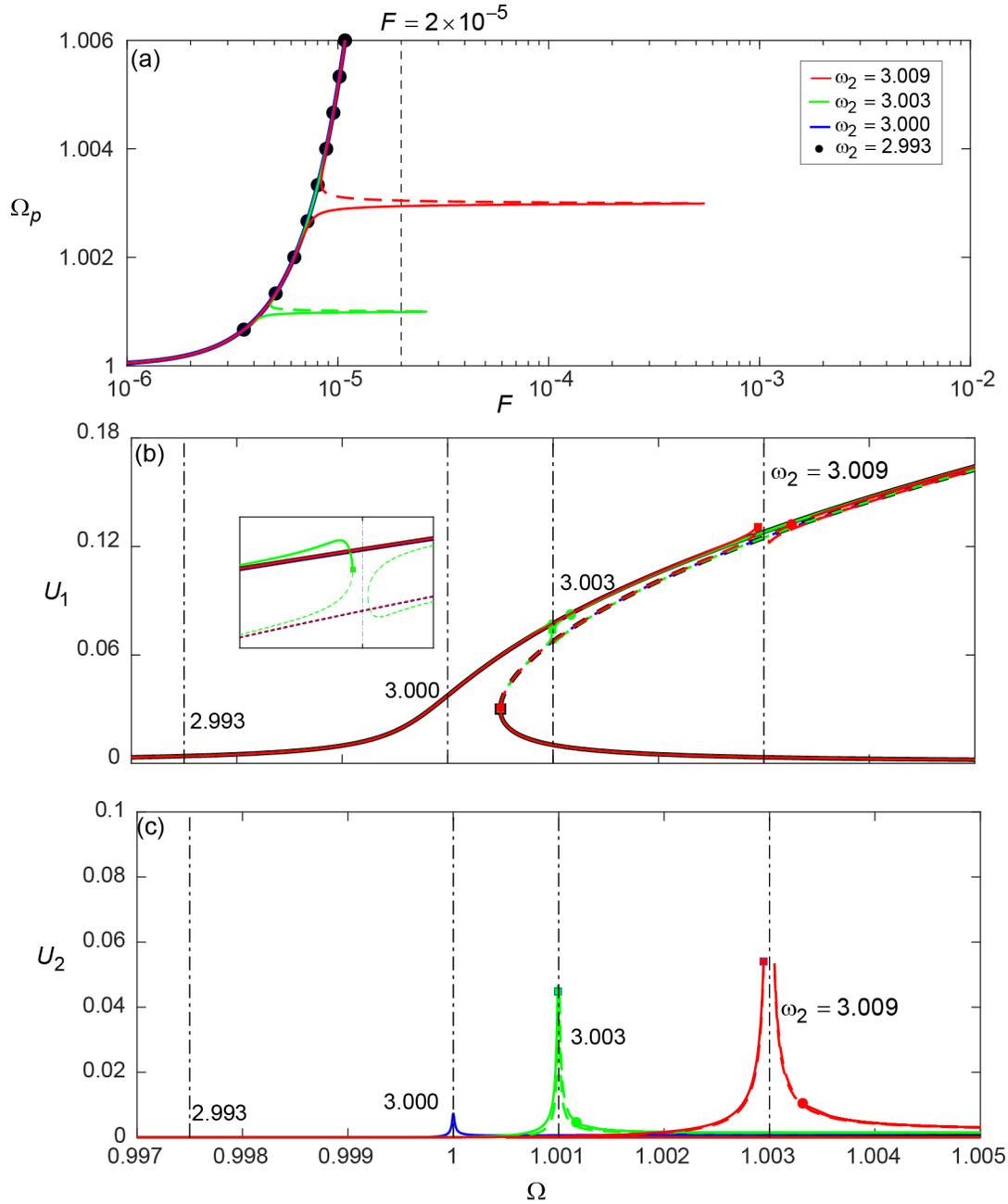

**Fig. 7** Influence of frequency mismatch on the forced response for 1:3 InRes at $\gamma_1 = 0.5, \alpha = 0.1$. (a) shows the peak frequency $\Omega_p$ versus forcing amplitude $F$ for $\omega_2 = [2.993, 3.000, 3.003, 3.009]$. (b) and (c) show the corresponding forced response curves of the first ($U_1$) and second modes ($U_2$), respectively, at $F = 2 \times 10^{-5}$. Dashed lines indicate unstable solutions. Symbols: (■) Saddle-Node (SN) bifurcations, and (●) Hopf bifurcations (HB).

## 4. Conclusions

This study provides a comprehensive theoretical investigation into the mechanisms of frequency stabilization in micromechanical resonators through 1:2 and 1:3 internal resonance (InRes). Using a generalized two-mode model that includes Duffing nonlinearity and nonlinear modal coupling, we examined how different parameters impact the frequency response and stabilization behavior of micromechanical resonators. Particularly, our analysis based on the $\pi/2$–backbone curve, FRCs, and sensitivity plots has proven to be a powerful approach: it not only enables the investigation of the steady-state forced response in open-loop operation but also facilitates the prediction of dynamic behavior for closed-loop operation, providing critical insights into practical operational conditions.

Our analysis identifies two distinct stabilization regimes based on the relative strength of coupling compared to the stiffening effect. For weak coupling, frequency stabilization is achieved through amplitude and frequency saturation over a range of forcing amplitudes for both 1:2 and 1:3 InRes. In this regime, the frequency becomes pinned to a narrow range, driven by the transfer of excess energy to the higher mode, which effectively suppresses the amplitude-frequency effect. In contrast, strong coupling leads to a more significant reduction in the amplitude-frequency effect directly in the $\pi/2$–backbone. For 1:2 InRes, this manifests as an asymptote in the $\pi/2$-backbone curve, gradually reducing the amplitude-frequency effect over an unconstrained range. Meanwhile, for 1:3 InRes, a zero-dispersion point emerges, directly eliminating the amplitude-frequency effect, but only over a limited range. These distinct stabilization mechanisms explain differences observed in previous experimental studies and theoretical predictions, highlighting the versatility and robustness of InRes in enhancing frequency stability.

Overall, this work provides valuable insights into the design and optimization of micromechanical resonators for improved frequency control. By identifying the conditions under which InRes effectively stabilizes frequency, this study underscores the potential of InRes as a robust tool for enhancing the performance of micromechanical resonators in time-keeping and sensing applications. The theoretical research presented here can be further validated by future experimental work, exploring the versatility in the design and operation of micromechanical resonators. Specifically, varying the coupling strength within a device—such as introducing asymmetry in microbeam resonators—can be achieved through pre-stress induced by electrostatic or thermal tuning. This approach will help further confirm and extend the findings of this study.

**Acknowledgement**

This work was financially supported in part by National Science Foundation (NSF-CMMI-2227527).